\documentclass[useAMS,usenatbib,amsmath,amssymb,floatfix]{mn2e}
\usepackage{amssymb}
\usepackage{graphicx}
\usepackage{bm}
\title[A new search for features in the primordial power spectrum]{A new search for features in the primordial power spectrum}

\author[Domenico Tocchini-Valentini, Marian Douspis and Joseph Silk]{Domenico Tocchini-Valentini\thanks{E-mail: dtv@astro.ox.ac.uk}, Marian Douspis\thanks{E-mail: douspis@astro.ox.ac.uk; Now at LATT-OMP, 14, av. E. Belin, F-31400 Toulouse, France.}
and Joseph Silk\thanks{E-mail: silk@astro.ox.ac.uk}\\
Department of Physics, University of Oxford,
Denys Wilkinson Building,\\
Keble Road, Oxford OX1 3RH, United Kingdomy}

\begin{document}

\date{Accepted ... Received ...; in original form ...}

\pagerange{\pageref{firstpage}--\pageref{lastpage}} \pubyear{2004}

\maketitle

\label{firstpage}

\begin{abstract}
We develop a new approach toward a high resolution non-parametric
reconstruction of the primordial power spectrum using WMAP cosmic
microwave background temperature anisotropies that we confront with
SDSS large-scale structure data in the range
$k\sim0.01-0.1\,h\mathrm{Mpc^{-1}}$. We utilise the standard
$\Lambda$CDM cosmological model but we allow the baryon fraction to
vary. In particular, for the concordance baryon fraction, we 
compare indications of a possible feature at
$k\sim0.05\,h\mathrm{Mpc^{-1}}$ in WMAP data with suggestions of similar features in large scale structure surveys.
\end{abstract}


\section{Introduction}

The Wilkinson Microwave Anisotropy Probe (WMAP) experiment has recently
provided a high resolution full sky cosmic microwave background (CMB)
map that provides
 an unprecedented opportunity to probe the structure
and the contents of the  universe (Spergel et al. 2003). The corresponding angular power spectrum has been used to set  constraints on cosmological parameters.
In the framework of $\Lambda$CDM cosmologies, such constraints are
strong if the initial conditions are parametrized by primordial power
spectra in the form of power laws with or without running spectral
index. However parameter estimation can be weakened and biased if more
complex forms of initial conditions are adopted, cf. \cite{dunkley} and \cite{bla}.
We explore another such possibility here.

Recent papers from the Sloan Digital Sky Survey (SDSS) team report
measurements of the matter power spectrum and the cosmological
parameters that give a best fit to its shape (Tegmark et al. 2003a,b; Pope et al. 2004).
%~\cite{sdss1,sdss2,pope},
In particular Pope et al. (2004) finds that for the 
for a value of the matter density, $\Omega_{m}h=0.264\pm0.043$ and
baryon fraction, $f_{b}=\Omega_{b}/\Omega_{m}=0.286\pm0.065$ at 1
$\sigma$. These determinations, in particular that of the baryon
density, are in marginal conflict with the data from both
WMAP (Spergel et al. 2003)
%~\cite{sper} 
and primordial nucleosynthesis (BBN)
%~\cite{bbn,cuo}
(Cyburt et al. 2003; Cuoco et al. 2003),
which, together with the Hubble constant
%~\cite{hst}
(Freedman et al. 2001) determination,
prefer a lower baryon fraction, $f_{b}\simeq0.2$ when a power law
primordial power spectrum in a $\Lambda$CDM cosmology is adopted.

Such a discrepancy {\it might} be due to the presence of intrinsic
features in the power spectrum that affect the SDSS (and other survey)
data, such as for example a dip and a bump at $k\sim0.035 \,h
\mathrm{Mpc}^{-1}$ and $k\sim0.05 \,h \mathrm{Mpc}^{-1}$ (see also Atrio-Barandela et al. 2001; Barriga et al. 2001). It is useful to explore the implications of such a possibility, even though the data at present is not compelling, because very similar features in the matter power spectrum have
been detected in independent large-scale structure surveys, notably
the 2dF, Abell/ACO cluster and PSCz surveys
%~\cite{einasto,2df,abel,pscz}.
(Einasto et al. 1997; Percival et al. 2001; Miller \& Batusky 2001; Hamilton et al. 2000).  
In this
work, we show that, if the power law hypothesis for the shape of
the initial power spectrum is relaxed, a still good or even better agreement between the
different data sets can be obtained. Our strategy is principally comprised
of two steps: a high definition reconstruction of the primordial power
spectrum from WMAP data (for given sets of cosmological parameters)
and a convolution of the initial power spectrum with the SDSS window
functions to determine the statistical agreement between the two data
sets.  We choose to study only these two data sets in a limited range
of $k-$space in order to make a preliminary exploration of the
possible influence of primordial features.  We find that for
reasonable choices of the cosmological parameters, the
WMAP-reconstructed matter power spectrum shows oscillations that are
very similar to those of SDSS once the convolution is performed. This
results in an improved concordance for the inferred baryon fraction
and at the same time is suggestive of a deviation from a power-law
primordial power spectrum.

\section[]{Method}

The temperature CMB angular power spectrum is related to the
primordial power spectrum, $P^{0}$, by a convolution with a window
function that depends on the cosmological parameters:
\begin{equation}
C_{\ell}=4\pi\int\Delta^{2}_{\ell}(k)\, P^0(k) \, \frac{\mathrm{d} k}{k} \simeq
WP^{0},
\end{equation}
where the last term is in matrix notation and represents a numerical
approximation to the integral, calculated using a modified version of
CMBFAST
%~\cite{cmbfast}.
(Seljak \& Zaldarriaga 1996).
The $C_{\ell}$ spectrum then contains information
about both the cosmological parameters and the initial power spectrum.

Various papers have presented reconstructions of the power spectrum
from WMAP data (e.g. 
%\cite{bridle,wang,shafi,kogo,matsu,hanne}).  
Bridle et al. 2003; Mukherjee \& Wang 2003; Matsumiya et al. 2003; Shafieloo \& Souradeep 2004; Kogo et al. 2004; Hannestad 2004).
Most of these have
confronted an \textit{a priori} parametrised power spectrum with the
data to obtain information about its shape and amplitude. Following
the spirit of \cite{gaw} and \cite{tegzald}, our power spectrum is \textit{not}
described by a small set of parameters that incorporate features, but
is finely discretized in $k$-space (see also Shafieloo \& Souradeep 2004; Kogo et al. 2004). Our work represents the first attempt to reconstruct the power spectrum at high resolution in the full range $k\sim0.01-0.1\,h\mathrm{Mpc^{-1}}$ together with an estimation of the error covariance matrix.
Our findings are consistent with the results obtained with a different method by \cite{kogo} that are limited to $k\lesssim0.045\,h\mathrm{Mpc^{-1}}$, corresponding to $\ell\lesssim430$. 

Generally speaking, it is not
possible to invert the matrix $W$ to solve for the
power spectrum, because each $C_{\ell}$ embraces information about a
limited range in $k$-space. However the inversion is feasible under
some assumptions. Basically the lack of information can be remedied by
the introduction of priors, such as for example the requirement of a
certain degree of smoothness in the solution. To be specific, we
consider a solution of the form 
\begin{equation}
\overline P=MC^{WMAP}_{\ell}
\end{equation}
with an error covariance matrix given by 
\begin{equation}
\Sigma=\left[ W^{t}N^{-1}W+\epsilon L^{t}L \right]^{-1}.
\end{equation}
Here: $M=\left[ W^{t}N^{-1}W+\epsilon L^{t}L \right]^{-1}W^{t}N^{-1}$;
$C^{WMAP}_{\ell}$ are the mean values of WMAP data; $N$ is the WMAP
covariance matrix (Verde et al. 2003) and $L$ is a discrete approximation to the first
derivative operator.  Our approach follows the spirit of a similar
solution proposed in \cite{tegzald}; however we replace the identity
operator by $L$.  The parameter $\epsilon$ regulates the degree of
solution smoothness, since the derivative operator $L$ acts in such a way to minimize the norm of the solution
derivative. This solution can be effectively thought of as a linear
least squares solution modified to accommodate smoothing.

In order to select a value for the parameter $\epsilon$, we calculate
the angular power spectrum $\overline C_{\ell}$ resulting from $\overline
P$. 
Then we form the expression 
\begin{equation}
\chi^{2}=(\overline C_{\ell}-C^{WMAP}_{\ell})^{t}N^{-1}(\overline C_{\ell}-C^{WMAP}_{\ell}).
\end{equation}
It is reasonable to fix the parameter $\epsilon$ to render the
$\chi^{2}$ in the previous equation equal to the number of WMAP
temperature data points, namely 899. In other words, a solution is
chosen that gives an acceptably good fit to the data.
% mod % 
Moreover, as will be explained below, this represents a conservative choice since the smoothing decreases the effective number of degrees of freedom.

To test the method, we reconstruct the power spectrum by using angular
power spectra calculated from various power spectrum shapes with and
without added errors. We obtain very good results, except when there
are features of extension comparable to our $k$-space gridding $\Delta
k$, which we took to be approximately $10^{-4}$ in the range $0.01-0.1
\,\mathrm{Mpc}^{-1}$. The integration range is
typically between $10^{-5}$ and $0.3\,\mathrm{Mpc}^{-1}$ and comprises
about 1500 points. We also verify that the signals detected in the
actual WMAP data (see below) do not depend on discretization issues by
testing different resolutions. Our method yields a $k$-space
resolution that is limited only by computing resources, at the price
however of introducing correlations between neighboring points that
we take fully into account into our statistical analysis.

In analogy with ordinary regression it is possible to define the effective number of degrees of freedom caused by smoothing by the expression $\mathrm{Trace}\left( WM \right)$. It is easy to recognize that without smoothing this quantity would have been equal to the number of bins in $k$-space. 
We have found that our choice of the parameter $\epsilon$ implies the effective usage of approximately 45 constraining informations out of the more than 1000 amplitudes that specify the power spectrum.
 
The procedure just described must be
repeated for each given set of cosmological parameters. We proceed by
first evolving the primordial power spectrum to redshift zero by
multiplying it by the appropriate transfer function calculated by
CMBFAST. Then we convolve the derived matter power spectrum with the
SDSS window functions given by \cite{sdss1}. Finally, we evaluate
$\chi^{2}$:
\begin{equation}
\chi^{2}=\left(\widetilde P-\frac{P^{SDSS}}{b^{2}}
\right)^{t}\left[\widetilde\Sigma+\frac{N^{SDSS}}{b^{4}}\right]^{-1}\left(
\widetilde P-\frac{P^{SDSS}}{b^{2}} \right), \label{cfr}
\end{equation} 
where $P^{SDSS}$ and $N^{SDSS}$ represent the SDSS mean data and
(diagonal) covariance matrix (Tegmark et al. 2003a), $\widetilde P$ and
$\widetilde \Sigma$ are the SDSS-convolved WMAP-reconstructed matter
power spectrum and covariance matrix (calculated by error propagation)
and $b$ is the bias parameter. We follow the common practice of
considering a constant bias which we set at the optimal
value. 

\section{Results}

In performing the $\chi^{2}$ analysis indicated by Eq.~(\ref{cfr}), we
restrict the computation to the SDSS experimental data corresponding
to the 11 points with lower $k$-values, such that
$k\lesssim 0.1\,h\mathrm{Mpc}^{-1}$, in order to avoid effects caused
by the limitation in $k$-range that arises since we are extracting
information from the CMB power spectrum and the quality of the WMAP data
degrades significantly above $\ell\sim 800$.

We select a few choices of the cosmological parameters, for which we
perform the reconstruction. The presence of features in the power
spectrum is associated with the deviations from a power law angular
power spectrum noticed by \cite{sper}. These authors mention that the
first release of data did not include effects contributing roughly
0.5-1\% to the power spectrum covariance, mainly due to gravitational
lensing, beam uncertainties and non-Gaussianity in the noise
maps. Therefore, to verify if they are real, a cross-check with an
independent dataset is needed. The SDSS data serve such a purpose.

In Table~\ref{table1}, we list the models used, defined in terms of
the baryonic and total matter content,
Hubble constant $h$ (in units of $\mathrm{100\,km\,sec^{-1}Mpc^{-1}}$)
and bias $b$. The corresponding $\chi^{2}$ for a selection the
first 11 SDSS data points with lowest wavenumber are also given
and will be used for the discussion below. The optical depth $\tau$ is
set equal to the best WMAP case, 0.166.
We have chosen to concentrate on a few cases that differ mainly with
respect to the baryonic fraction, since this is the most interesting
quantity to probe in light of the SDSS analysis that uses a power-law
power spectrum. 
\begin{table}
\caption{\label{table1} Models, inferred optimal bias,  and  $\chi^{2}$
comparison between WMAP constrained matter power spectra and the SDSS
data for a selection of the first 11 points with lowest
wavenumber, corresponding to 10 degrees of freedom.}
\begin{tabular}{cccccccc} \hline
 Model&$f_{b}$&$\Omega_{m} h$&$\omega_b$&$h$&$b$&$\chi^{2}/dof$\\ \hline
1& 0.171 & 0.194 & 0.024  & 0.72 & 1.04 & 4.7/10   \\ 
2& 0.286 & 0.264 & 0.054  & 0.72 & 0.97 & 26.4/10 \\ 
3& 0.155 & 0.215 & 0.023  & 0.695 & 1.10 & 3.3/10  \\ 
4& 0.200 & 0.209 & 0.030  & 0.72 & 1.11 & 3.6/10  \\ 
5& 0.143 & 0.194 & 0.020  & 0.72 & 1.00 & 6.4/10  \\ \hline
\end{tabular}
\end{table}

We label the models as: model 1, the WMAP mean best case (Spergel et al. 2003);
model 2, the SDSS best case (Pope et al. 2004) (in marginal conflict with the
previous one); model 3, close to the joint WMAP-SDSS best case (Tegmark et al. 2003b);
model 4 and model 5, characterized by a relatively high and low
baryonic content. The ``best cases'' we refer to were all obtained for
a power-law power spectrum.  The latest constraints from primordial
nucleosynthesis give, at 1 $\sigma$:
$\omega_{b}=h^{2}\Omega_{b}=0.022\pm0.002$ (Cyburt et al. 2003; Cuoco et al. 2003), which excludes 
model 2 with $\omega_{b}=0.05$ .

\begin{figure}
\includegraphics[totalheight=0.19\textheight,viewport=110 150 200 200,clip]{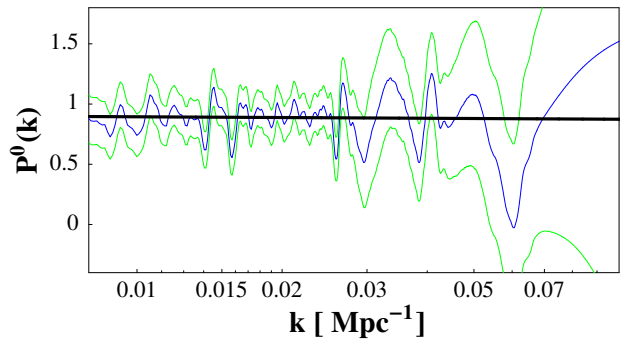}
\caption{\label{fig.1} Reconstructed initial power spectrum in units
  of $2.95\times 10^{-9}$ from WMAP $C_\ell$'s with model 3
  cosmological parameters. The irregular middle curve represent the
  mean and the surrounding curves are obtained by summing and
  subtracting the square root of the diagonal elements of the
  covariance matrix. 
The smooth line is a reference
power law power spectrum given by 
 $k^{n_{s}-1}$, where $n_{s}\approx 0.99$.}
\end{figure}

\begin{figure}
\includegraphics[totalheight=0.19\textheight,viewport=110 150 200 200,clip]{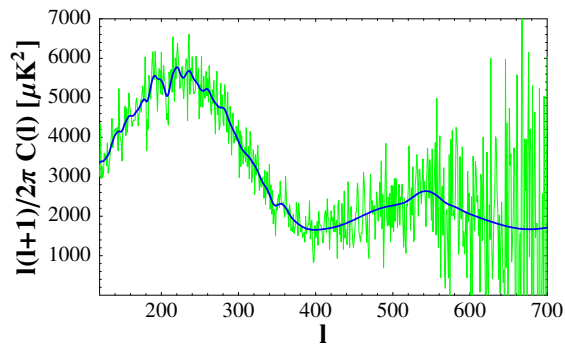}
\caption{\label{fig.2} The noisy thin curve connects the data from WMAP. The reconstructed power spectrum convolved back to multipole space is given by the thick smooth line, that shows clearly the effect of smoothing.}
\end{figure}

Our first result is a reconstruction of  the initial
power spectrum shape from the CMB angular power spectrum with the associated
error covariance matrix. Fig.~\ref{fig.1} shows the corresponding initial spectrum
reconstructed from the WMAP data and
assuming the cosmological parameters of model 3, in the range $0.01 <
k < 0.075\,\mathrm{Mpc}^{-1}$. The effective range probed in $\ell$ is approximately 140-800. By reconvolving the power spectrum back to multipole space, it is possible to note in Fig.~\ref{fig.2} the action of smoothing and appreciate the accuracy of the inversion. Even if not far from a power law, small features are present in the initial power spectrum at all scales; we shall devote our attention mostly to the succession of bumps and dips around $0.035\,\mathrm{Mpc}^{-1}$. 

\begin{figure}
\includegraphics[totalheight=0.19\textheight,viewport=110 150 200 200,clip]{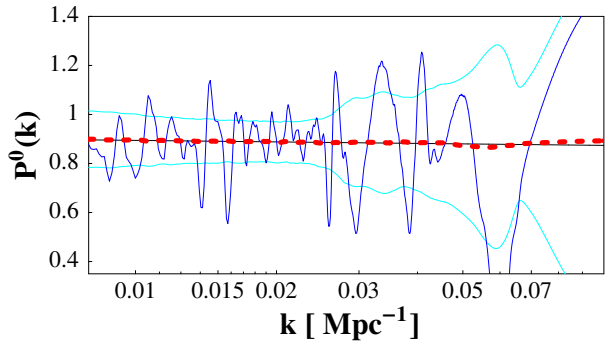}
\caption{\label{fig.3} Results from 5000 Monte Carlo realizations drawn form a power law power spectrum, given by 
 $k^{n_{s}-1}$ where $n_{s}=0.99$, with model 3
  cosmological parameters. The thin smooth line in the center is the exact initial power law and the dashed curve is the averaged reconstructed initial power spectrum. The two symmetric lines represent the standard deviation from the averaged reconstruction. No strong systematic bias appears in the reconstruction and possible candidates for features are given by the outliers respect to the one sigma band. The plot is in units of $2.95\times 10^{-9}$.}
\end{figure}

It is useful to implement Monte Carlo simulations to look for deviations and systematic biases in the reconstruction. We generated 5000 random realizations around the angular power spectrum calculated from a power law primordial power spectrum assuming a gaussian distribution, which is reasonable in the probed multipole range. The case for model 3 is presented in Fig.~\ref{fig.3}. The averaged reconstruction doesn't show strong bias and deviations bigger that the one sigma band, obtained from the standard deviation respect to the averaged  reconstruction, are also evident. Again, there are indications of possible features at $k\sim0.035\, \mathrm{Mpc^{-1}}$ and other wave numbers.

\begin{figure}
\includegraphics[totalheight=0.35\textheight,viewport=0 0 200
200,clip]{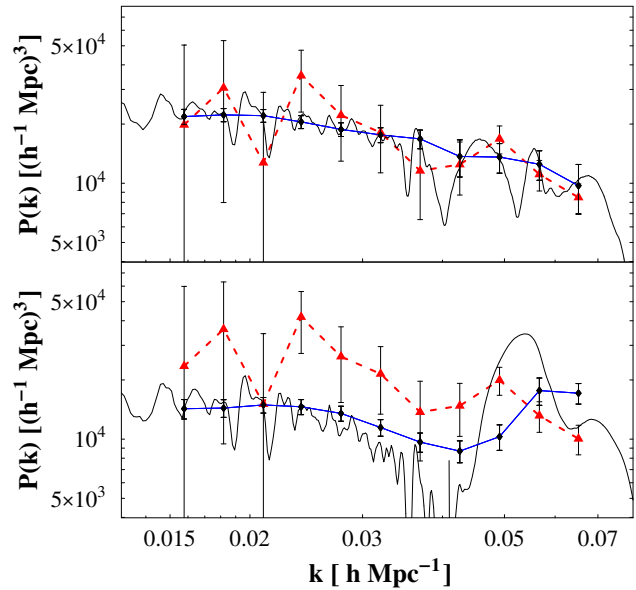}
\caption{\label{fig.4} The upper plot shows the WMAP best case (model 1). 
The reconstructed matter power spectrum convolved with the SDSS window
functions and evolved at redshift zero is given by the black diamonds
connected by a piecewise continuous line. The small error bars reflect
the quality of WMAP data and are constructed with the square root of
the diagonal elements of the full covariance matrix, which is
used for the statistical analysis. The thin irregular curve is the
mean reconstruction before convolution and the dash-joined triangles
are the SDSS data with a maximising bias. The lower plot is the SDSS
best case (model 2)}
\end{figure}

Models 1 and 2 reconstructed matter power spectra are plotted in
Fig.~\ref{fig.4}.
Model 2, constructed to fit the SDSS data, is marginally excluded by
 WMAP data under the initial power law assumption: it requires too
 many baryons in order to fit the features (oscillations) of the SDSS
 reconstructed matter power spectrum. If we relax this assumption by
 using our method and keeping the same baryon fraction, we see that
 the WMAP $C_\ell$ features translated into k-space do not match the SDSS
 features ($\chi^2=26/10$). Baryonic oscillations and CMB features do
 not converge to reproduce the SDSS behaviour. As noted in \cite{pope},
 oscillations induced by the prefered baryon fraction value of CMB are
 out of phase with SDSS ``oscillations''.

Conversely, assuming a power law initial power spectrum, model 1
gives a poor fit to CMB data (Spergel et al. 2003) and a good fit to the
selected SDSS data ($\chi^{2}=6/10$). When we allow for a free shape for the
initial power spectrum, using our method, model 1 becomes by
definition a good fit to WMAP $C_\ell$'s and decently fits
the 11 points selected from SDSS ($\chi^{2}=5/10$). By taking into
account the features in the CMB we improve not only the fit to the
CMB but also to the SDSS data. 

\begin{figure}
\raggedleft \includegraphics[totalheight=0.46\textheight,viewport=0 0
155 200,clip]{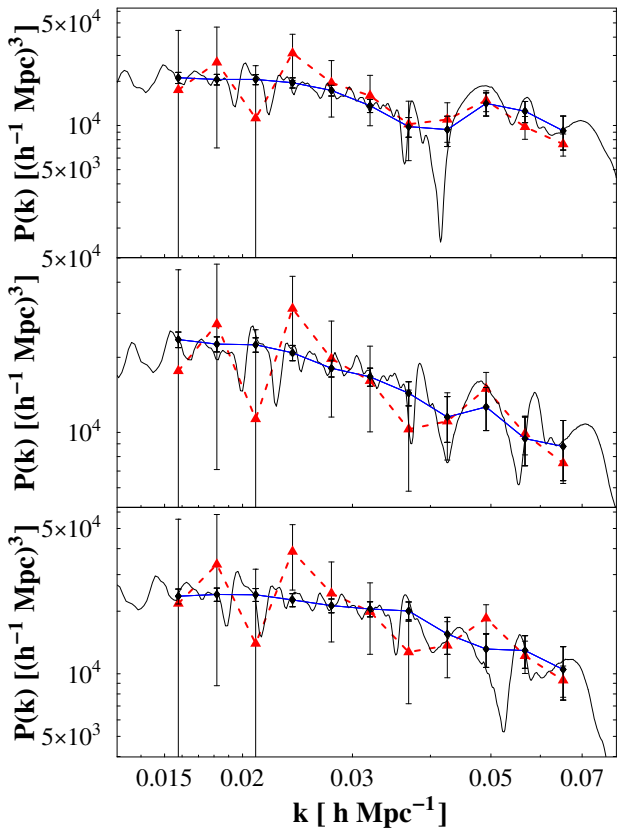}
\caption{\label{fig.5} The conventions are the same as for the
previous figure. In the lower plot, model 5, with a low baryonic
content, presents oscillations completely out of phase with some of
the SDSS features. In the middle, model 3, our best case, mimics
particularly well the SDSS feature at
$k\sim0.05\,h\mathrm{Mpc^{-1}}$. Model 4 at the top represents a case
with slightly higher baryonic fraction and is also a good fit.}
\end{figure}

However we found other models that can achieve a better fit. If we consider the cosmological parameter of model 3 (close to the best
WMAP+SDSS fit), our reconstructed matter power spectrum gives a better
fit ($\chi^{2}=3.3/10$, see Fig.~\ref{fig.5}). In doing this, we show
that the same model could fit both the WMAP $C_\ell$'s and SDSS matter
power spectrum shape.  This last point is our principal result, and
suggests that an acceptable compromise that simultaneously fits both
WMAP and SDSS may require a change respect to the concordance model, such as
\textit{deviations from a power-law initial power spectrum}. 

We then investigated the effect of varying the baryonic fraction in
model 3 and find that features from the CMB angular spectrum must
be combined with the appropriate  amount of baryons in order to fit the SDSS
``oscillations'' (see Fig.~\ref{fig.5}). Model 5, with a low baryonic content, is a relatively bad fit because the WMAP features seem almost out of phase with the SDSS features. While model 4 leads to a better agreement since the shapes of the SDSS $k\sim0.035$ and
$0.05\,h\mathrm{Mpc^{-1}}$ features are mimicked reasonably well. We furthermore  note
that model 3 falls in the observational range of both $h$
($0.72\pm0.07$, Freedman et al. 2003) and $\Omega_bh^2$ ($0.022\pm0.002$,
Cyburt et al. 2003; Cuoco et al. 2003). 

\begin{figure}
\includegraphics[totalheight=0.2\textheight,viewport=110 160 200
210,clip=]{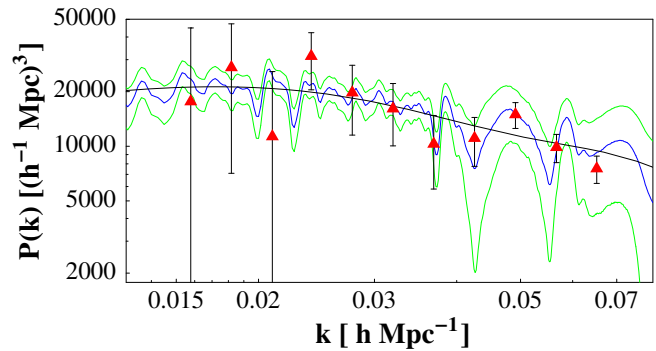}
\caption{\label{fig.6} The mean (central curve) of our
best case reconstructed matter power spectrum, model 3, is shown.  The
surrounding curves are obtained by summing and subtracting the square
root of the diagonal elements of the covariance matrix. This is
strictly legitimate only for an uncorrelated case, however here it is
used to  convey the magnitude of the uncertainties. The thick line is
calculated from a primordial power spectrum given by $k^{n_{s}-1}$, where $n_{s}\approx 0.99$.
Most of the detected features do not seem to be the product of baryonic oscillations (for a possible explanation see e.g. Martin \& Ringeval 2004). For convenience, the SDSS data are also depicted.}
\end{figure}

Finally, looking at Fig.~\ref{fig.6} suggests that the feature seen at
$k \sim 0.05\, h \rm \mathrm{Mpc}^{-1}$ in the SDSS is well reproduced by the
reconstructed WMAP data with model 3 cosmological parameters. 

% mod 
More precisely, taking into account the effective number of degrees of freedom defined above, it is feasible to compare the probability of model 3 with the power law case, considering both the WMAP and SDSS data. For a power law the probability is 3.9\%, considering as free parameters $h$, $\Omega_{b}$, $\Omega_{c}$, $\tau$, $b$, $n_{s}$ and the amplitude of the fluctuations. While for model 3, with free parameters $h$, $\Omega_{b}$, $\Omega_{c}$, $\tau$, $b$ and the effective contribution from the smoothed amplitudes that amounts to 45.2, the probability is 11.9\%.

\begin{figure}
\includegraphics[totalheight=0.2\textheight,viewport=110 160 200
210,clip=]{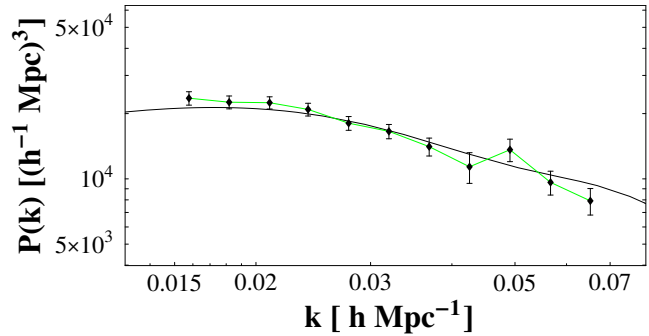}
\caption{\label{fig.7} The
 power spectrum produced from a primordial power law with $k^{n_{s}-1}$, where $n_{s}\approx 0.99$ (smooth line) is depicted together with the combined spectrum from the SDSS data (data points with error bars) and the SDSS-convolved reconstruction for model 3 (see text). The SDSS feature at $k\sim0.05\,h\mathrm{Mpc^{-1}}$ appears to be present at 1 sigma c.l..}
\end{figure}

In addition to a global goodness of fit, a statistical test to search for local features can be implemented. We combined the SDSS data and the SDSS-convolved WMAP-reconstructed power spectrum for model 3 to show more clearly where it is more likely to have deviations from a power law. The resulting power spectrum is depicted in Fig.~\ref{fig.7}
and its error covariance matrix and mean are given by
\begin{equation}
\Sigma_{comb}=\left[\widetilde\Sigma^{-1}+\left(\frac{N^{SDSS}}{b^{4}}\right)^{-1}\right]^{-1}
\end{equation}
\begin{equation}
P_{comb}=\Sigma_{comb}\left[\widetilde\Sigma^{-1}\widetilde P+\left(\frac{N^{SDSS}}{b^{4}}\right)^{-1}\frac{P^{SDSS}}{b^{2}}\right].
\end{equation}
 The feature at $k\sim0.05\,h\mathrm{Mpc^{-1}}$ stands out as a deviation from a power law at a significance of %approximately
 one sigma. Also note that the interesting eighth and ninth points are positively correlated at a level of approximately 20\%. Similar detections in other surveys suggest that this feature may be real and certainly encourage further searches in future experiments. 

The resolution of the released SDSS data does not allow us to test the
reality of various fine features that seem to be present in the
reconstructed power spectrum. Along with progress in large scale structure measurements, a detailed understanding in forthcoming CMB experiments of possible systematics in beam uncertainties and eventual non-gaussianities coming from foregrounds and point-sources residues would definitely increase the chances of detection of features in the primordial power spectrum.

\section{Discussion}

In summary we have shown that a slight discrepancy in the baryonic
fraction that arises from SDSS and WMAP data could be resolved if the
primordial power spectrum is not fixed \textit{a priori} in the form
of a power law, but is constrained to satisfy CMB data when confronted
with large-scale structure information and letting key cosmological
parameters vary. Our method is feasible due to a high resolution
reconstruction of the power spectrum from the CMB angular power
spectrum, that is carried out for the first time in the full range considered, and provides evidence in favour of intrinsic features in the
primordial power spectrum. In particular a feature at
$k\sim0.05\,h\mathrm{Mpc^{-1}}$ seems to occur in both the WMAP and
SDSS data sets. This kind of technique offers an effective way to
break the degeneracy between the baryon abundance and the shape of the
power spectrum (in the conventional approach parametrized by the
spectral index, $n_{s}$) and relevantly brings also to a measure of the galaxy bias.
 
A power spectrum with the observed features brings to a good fit of SDSS data without the necessity of a large baryon fraction. More generally, if the initial power law hypothesis is relaxed, deviations seem to
induce a somewhat better concordance picture from the combination of
CMB, large-scale structure, BBN and Hubble constant measurements. This
could have important consequences for fundamental physics.

\section*{Acknowledgments}
DTV would like to thank Y. Hoffman, P. Ferreira, C. Skordis and K. Moodley for
useful discussions. DTV acknowledges a Scatcherd Scholarship. MD
acknowledges financial support provided through the
 E. U. Human Potential Program under contract
 HPRN-CT-2002-00124, 
CMBNET. 
%\bibliography{projectpk}

\bsp

\label{lastpage}

\end{document}